# Karl Pearson's Theoretical Errors and the Advances They Inspired

**Stephen M. Stigler**

*Abstract.* Karl Pearson played an enormous role in determining the content and organization of statistical research in his day, through his research, his teaching, his establishment of laboratories, and his initiation of a vast publishing program. His technical contributions had initially and continue today to have a profound impact upon the work of both applied and theoretical statisticians, partly through their inadequately acknowledged influence upon Ronald A. Fisher. Particular attention is drawn to two of Pearson's major errors that nonetheless have left a positive and lasting impression upon the statistical world.

## 1. INTRODUCTION

Karl Pearson surely ranks among the more productive and intellectually energetic scholars in history. He cannot match the most prolific humanists, such as one of whom it has been said, "he had no unpublished thought," but in the domain of quantitative science Pearson has no serious rival. Even the immensely prolific Leonhard Euler, whose collected works are still being published more than two centuries after his death, falls short of Pearson in sheer volume. A list of Pearson's works fills a hardbound book; that book lists 648 works and is still incomplete (Morant, 1939). My own moderate collection of his works—itself very far from complete (it omits his contributions to *Biometrika*)—occupies 5 feet of shelf space. And his were not casually constructed works: when a student or a new co-worker would do the laborious calculations for some statistical analysis, Pearson would redo the work to greater accuracy, as a check. An American visiting Pearson in the early 1930s once asked him how he found the time to write so much and compute so much. Pearson replied, "You Americans would not understand, but I never answer a telephone or attend a committee meeting" (Stouffer, 1958).

Pearson's accomplishments were not merely voluminous; they could be luminously enlightening as well. Today the most famous of these are Pearson's Product Moment Correlation Coefficient and the Chi-square test, dating respectively from 1896 and 1900 (Pearson, 1896, 1900a, 1900b). He was a driving force behind the founding of *Biometrika*, which he edited for 36 years and made into the first important journal in mathematical statistics. He also established another journal (the *Annals of Eugenics*) and several additional serial publications, two research laboratories, and a school of statistical thought. Pearson pioneered in the use of machine calculation, and he supervised the calculation of a series of mathematical tables that influenced statistical practice for decades. He made other discoveries, less commonly associated with his name. He was in 1897 the first to name the phenomenon of "spurious correlation," thus publicly identifying a powerful idea that made him and countless descendents more aware of the pitfalls expected in any serious statistical investigation of society (Pearson, 1897). And in a series of investigations of craniometry he introduced the idea of landmarks to the statistical study of shapes.

*Stephen M. Stigler is the Ernest DeWitt Burton Distinguished Service Professor in the Department of Statistics, University of Chicago, 5734 University Avenue, Chicago, Illinois 60637, USA e-mail: stigler@uchicago.edu. This paper is based upon a talk presented at the Royal Statistical Society in March 2007, at a symposium celebrating the 150th anniversary of Karl Pearson's birth.*

Pearson was at one time well known for the Pearson Family of Frequency Curves. That family is seldom referred to today, but there is a small fact (really a striking discovery) he found in its early development that I would call attention to. When we think of the normal approximation to the binomial, we usually think in terms of large samples. Pearson discovered that there is a sense in which the two distributions agree exactly for even the smallest number of trials. It is well known that the normal density is characterized by the differential equation

$$\frac{d}{dx}\log(f(x)) = \frac{f'(x)}{f(x)} = -\frac{(x-\mu)}{\sigma^2}.$$

Pearson discovered that $p(k)$, the probability function for the symmetric binomial distribution ($n$ independent trials, $p = 0.5$ each trial), satisfies the analogous difference equation exactly:

$$\frac{p(k+1) - p(k)}{(p(k+1) + p(k))/2} = -\frac{(k+1/2) - n/2}{(n+1) \cdot 1/2 \cdot 1/2}$$

or

$$\frac{\text{rate of change } p(k) \text{ to } p(k+1)}{\text{average of } p(k) \text{ and } p(k+1)}$$
$$= -\frac{\text{midpoint of } (k, k+1) - \mu_n}{\sigma_{n+1}^2}$$

for all $n$, $k$. The appearance of $n+1$ instead of $n$ in the denominator might be considered a minor fudge, but the equation still demonstrates a really fundamental agreement in the shapes of the two distributions that does not rely upon asymptotics (Pearson, 1895, page 356).

All of these Pearsonian achievements are indeed substantial, and constitute ample reason to celebrate him 150 years after his birth. But if these are all we saluted, I would hold that Pearson is being underappreciated. To properly gauge his impact upon modern statistics, we must take a look at parts of two works of his that are typically not held in high regard. Indeed, they are usually mentioned in derision, as exhibiting two major errors that show Pearson's limitations and highlight the great gulf that lay between Pearson and the Fisherian era that was to follow. I wish to return now to these two works and reassess them. I intend to argue that these errors should count among the more influential of his works, and that they helped pave the way for the creation of modern mathematical statistics.

## 2. PEARSON'S FIRST MAJOR ERROR

Louis Napoleon George Filon was born in France, but his family moved to England when he was three years old (Jeffrey, 1938). He first encountered Karl Pearson as a student at University College London. After receiving a B.A. in 1896, Filon served as Pearson's Demonstrator until 1898, and together they wrote a monumental memoir on the "probable errors of frequency constants," a paper read to the Royal Society in 1897 and published in their *Transactions* in 1898. In 1912 Filon succeeded Pearson as Goldsmid Professor of Applied Mathematics and Mechanics. At Pearson's retirement banquet in 1934, Filon (who was by that time Vice-Chancellor of the University of London) explained the genesis of this, their only work together in statistics.

> "K. P. lectured to us on the Mathematical Theory of Statistics, and on one occasion wrote down a certain integral as zero, which it should have been on accepted principle. Unfortunately I have always been one of those wrong-headed persons who refuse to accept the statements of Professors, unless I can justify them for myself. After much labour, I actually arrived at the value of the integral directly— and it was nothing like zero. I took this result to K. P., and then, if I may say so, the fun began. The battle lasted, I think, about a week, but in the end I succeeded in convincing Professor Pearson. It was typical of K. P. that, the moment he was really convinced, he saw the full consequences of the result, proceeded at once to build up a new theory (which involved scrapping some previously published results) and generously associated me with himself in the resulting paper" (Filon, 1934).

The term "probable error" was introduced early in the 19th century to mean what we would now call the median error of an estimate. Thus it is a value which, when divided by 0.6745, gives the standard deviation for an unbiased estimate with an approximately normal distribution. My guess is that the lecture that Filon referred to involved the formula for the probable error of Pearson's product-moment estimate of the correlation coefficient for bivariate normal distributions. Pearson had given this incorrectly in 1896, and one of the signal achievements of the Pearson–Filon paper was to correct that error (Stigler, 1986, page 343). But the 1898 paper did much more: it purported to give the approximate distributions for the probable errors of all the estimated frequency constants, indeed their entire joint distribution, for virtually any statistical problem. The theory presented was relatively short; most of the paper was taken up with a large number of applications.

Unfortunately, quite a number of these applications proved to be in error. There are some indications Pearson may have realized this by 1903, but if he did sense trouble with the paper, he did not call it to public attention. In 1922 Ronald Fisher repaired Pearson's omission when he noted in particular that outside of the case of the normal distribution, nearly all of the applications in Pearson–Filon were erroneous. This included many method-of-moments estimates (the gold-standard method for the Pearsonian school). A significant Pearson achievement came to be labeled an error, one eventually overcome by a Fisher success, and in consequence, the 1898 paper has suffered a low reputation. But despite these problems, the paper had, arguably, a significant and largely underappreciated positive impact upon statistics.

At first glance the Pearson–Filon argument may appear strikingly modern, apparently expanding a log-likelihood ratio to derive an asymptotic approximately multivariate normal distribution for the errors of estimation. The authors considered a multivariate set of $m$-dimensional measurements $x_1, x_2, x_3, \ldots, x_m$ of "a complex of organs," and they stated the "frequency surface" should be given by

"$z = f(x_1, x_2, x_3, \ldots x_m; c_1, c_2, c_3, \ldots c_p)$,

where $c_1, c_2, c_3, \ldots, c_p$, are $p$ frequency constants, which define the form as distinguished from the position of the frequency surface, and which will be functions of standard deviations, moments, skewnesses, coefficients of correlation, &c., &c., of individual organs, and of pairs of organs in the complex."

The "position" of the surface would be given in terms of the means $h_1, h_2, \ldots, h_m$ of the $x$'s, which are implicit in this notation as they are stated to give the origin of the surface, and so there are $m + p$ frequency constants to be determined from a set of $n$ measurements, each $m$-dimensional.

To determine the "probable errors" of the frequency constants, Pearson and Filon looked at the ratio formed by dividing the product of $n$ such surfaces (for the $n$ vector measurements) into what a similar product would be if the frequency constants had been different values:

$$\frac{P_\Delta}{P_0} = \Pi f(x_1 + \Delta h_1, x_2 + \Delta h_2, \ldots, x_m + \Delta h_m;$$
$$c_1 + \Delta c_1, c_2 + \Delta c_2, \ldots, c_p + \Delta c_p)$$
$$/\Pi f(x_1, x_2, \ldots, x_m; c_1, c_2, \ldots, c_p).$$

The logarithm of this ratio is then the difference of two sums. After being expanded "by Taylor's theorem," this yields a series with typical terms, writing $S$ for summation, they found to be

$$\log(P_\Delta/P_0)$$
$$= \Delta h_r S \frac{d}{dx_r}(\log f) + \frac{1}{2}(\Delta h_r)^2 S \frac{d^2}{dx_r^2}(\log f)$$
$$+ \Delta h_r \Delta h_{r'} S \frac{d^2}{dx_r\, dx_{r'}}(\log f)$$
$$+ \Delta c_s S \frac{d}{dc_s}(\log f) + \frac{1}{2}(\Delta c_s)^2 S \frac{d^2}{dc_s^2}(\log f)$$
$$+ \Delta c_s \Delta c_{s'} S \frac{d^2}{dc_s\, dc_{s'}}(\log f)$$
$$+ \Delta h_r \Delta c_s S \frac{d^2}{dh_r\, dc_s}(\log f) + \cdots$$
$$+ \text{cubic terms in } \Delta h \text{ and } \Delta c + \&c,$$

"where $f$ stands for $f(x_1, x_2, x_3, \ldots, x_m;\ c_1, c_2, c_3, \ldots, c_p)$."

Pearson and Filon then "replace sums by integrals," for example, by replacing the second summation above, namely $S\frac{d^2}{dx_r^2}(\log f)$, by $-B_r = \iiint \cdots f \frac{d^2 \log f}{dx_r^2}\, dx_1\, dx_2 \cdots dx_m$. With their notation the frequency surface encompassed a volume $= n$ (i.e., it was not a *relative* frequency surface), so if the $f$ were taken as a density or relative frequency surface this would be tantamount to replacing $\frac{1}{n} S \frac{d^2}{dx_r^2}(\log f)$ by its expectation $E[\frac{d^2}{dx_r^2}(\log f)]$, which would equal

$-B_r$. The integrals were then evaluated, and higher-order terms discarded, to get

$$P_\Delta = P_0 \exp\mathrm{t}. -\tfrac{1}{2}\{B_r(\Delta h_r)^2 - 2C_{rr'}\Delta h_r \Delta h_{r'}$$
(1)
$$- 2G_{rs}\Delta h_r \Delta c_s + E_s(\Delta c_s)^2$$
$$- 2F_{ss'}\Delta c_s \Delta c_{s'} + \&c. \cdots\}$$

where $B_r$, etc. are integrals given in terms of derivatives of $\log f$. We are then told:

> "This represents the probability of the observed unit, i.e. the individuals ($x_1$, $x_2$, $x_3,\ldots x_\mathrm{m}$, for all sets), occurring, on the assumption that the errors $\Delta\mathrm{h}_1$, $\Delta\mathrm{h}_2,\ldots$, $\Delta\mathrm{h}_\mathrm{m}$, $\Delta c_1,\ldots,\Delta c_\mathrm{p}$, have been made in the determination of the frequency constants. In other words, we have here the frequency distribution for errors in the values of the frequency constants."

Several of their steps, such as the cavalier substitution of integrals for sums or the discarding of remainder terms, may seem insufficiently defended, but the general drift is so similar to what we tend to see today that it would be easy for an uncritical reader to accept it, believing that it is probably essentially accurate, and that with some effort and additional regularity conditions all should be well. After all, such a reader might say, the replacement can be justified under reasonable regularity conditions, and even the last step would be sanctioned by a loose inverse probability argument such as was common at that time. That reader would be wrong.

## 3. THE SOURCE OF THE ERROR

The key to understanding what went wrong with the Pearson–Filon argument is at the very beginning, as a closer reading shows. Modern readers have understandably tended to take the frequency surface $z = f(x_1, x_2, x_3, \ldots, x_m; c_1, c_2, c_3, \ldots, c_p)$ as a parametric model. But in fact, in their notation, $z$ is the *fitted* surface in terms of *estimated* $h$'s and $c$'s. Pearson and Filon explained this implicitly in the first paragraph on page 231, when they write exclusively in terms of the group of $n$ individuals and the "means" $h_i$ (referring to the arithmetic means) as determining the origin of the surface, and explicitly in the short fourth paragraph, where they refer to the $h+\Delta h$'s and $c+\Delta c$'s of the numerator as being considered "instead of the observed values." This runs counter to modern statistical practice, which would focus on a specification in terms of parameters, not the estimates, with one value considered the true or population value, and the focus would be the deviations of the estimates from that value. The idea of this type of parametric modeling was, however, only to be introduced in 1922 by Fisher (Stigler, 2005), and Pearson's "frequency constants" were not parameters, even if they were sometimes employed in an equivalent fashion. This difference was, as we shall see, highly consequential; it was the source of the principal difficulties in the argument.

Because Pearson and Filon took the estimates as a starting point, the Taylor expansion they gave was about the *estimated* values. The expansion itself is fine, but when they came to substituting integrals for sums, they inadvertently encountered a problem. Consider the first sum that involves a general frequency constant $c$, namely $S\frac{d}{dc_s}(\log f)$. (The earlier terms involve the $h$'s but have been written in terms of derivatives with respect to the $x$'s; the issue is more clearly addressed with the $c$'s.) If we took $f$ as a density, it might not be unreasonable to replace this sum (divided by $n$) by its limit in probability, the expectation $E[\frac{d}{dc_s}(\log f)]$. But under what distribution should the expectation be computed? With Fisher it would be computed under the distribution with the true values of the parameters. But Pearson lacked that notion; for him there was no "true value," only a summary estimate in terms of observed values.

Someone—perhaps Fourier—has been quoted as saying that "Mathematics has no symbols for confused ideas." Anyone seeking a counterexample to this need look no further than Pearson–Filon. With their symbol $f = f(x_1, x_2, x_3, \ldots, x_m; c_1, c_2, c_3, \ldots, c_p)$ submerging the role of estimated values and elevating them in the process to surrogates for true values, the argument goes astray. All expectations are computed as if the estimated values were true values, and the result is a distribution for errors that does not in any way depend upon the method used to estimate. Pearson and Filon replaced $S\frac{d}{dc_s}(\log f)$ by the integral $D_s = \iiint \cdots f\frac{d\log f}{dc_s}dx_1\,dx_2\cdots dx_m$, which reduces identically to zero (if the two $f$'s are identical) under fairly general regularity conditions. But the same would not be generally true for $\iiint \cdots f(x|\theta)\frac{d\log f(x|\hat\theta)}{dc_s}\,dx_1\,dx_2\cdots dx_m$, writing $\hat\theta$ for $(h_1, h_2, h_3,\ldots,h_m; c_1, c_2, c_3,\ldots,c_p)$, and $\theta$ for the potential true value $\hat\theta$ is intended to estimate.

The summation $S\frac{d}{dc_s}(\log f)$ is identically zero if maximum likelihood estimates are used, but it remained for Fisher to notice that this term is not in general negligible; in fact, it will contribute asymptotically to the variance term if the estimate $\hat{\theta}$ is inefficient, as would be the case for many of Pearson's moment-based estimates. Pearson and Filon wrote that the expansion represented the distribution "on the assumption that the errors $\Delta h_1, \Delta h_2, \ldots, \Delta h_m$, $\Delta c_1, \ldots, \Delta c_p$ have been made in the determination of the frequency constants." But that is not what they had done. With their notation of a simple $f$ without arguments, and their fixation on the estimated values, they were led to mathematical misadventure.

If the substitution of limiting integrals for sums had been valid, the likelihood ratio $P_\Delta/P_0$ would then have been the ratio of the probability densities of the sample with estimated $h$'s and $c$'s (the denominator) to that for a hypothetical set of alternative values (the $h + \Delta h$'s and $c + \Delta c$'s, the numerator). In modern terminology, Pearson's final expression (1) was claimed to be an approximation for $P_\Delta$, the (conditional) density of the sample $x$ given that the estimated values are in error by $\Delta h$ or $\Delta c$, and their final claim ("In other words, we have here the frequency distribution for errors in the values of the frequency constants") was an assertion that formula (1) also gives the (conditional) density of the errors $\Delta h$ and $\Delta c$ given the data $x$. This last statement was not explained, but later in 1916 correspondence with Fisher (quoted in Stigler, 2005) Pearson described it as the use of inverse probability, making it a naive Bayesian approach with a uniform prior, such as was practiced routinely over the 19th century and was sometimes referred to as the Gaussian method.

Pearson's contemporaries did not raise questions about the memoir. When Edgeworth discussed and extended it in 1908, he gave no indication he saw anything amiss (Edgeworth, 1908). Only in 1922 did Ronald Fisher criticize the approach of the paper in a lengthy footnote (Fisher, 1922a, page 329), writing that "It is unfortunate that in this memoir no sufficient distinction is drawn between the *population* and the *sample* ...." Fisher went on to say that the results implicitly assume the estimates actually maximized the likelihood function, whereas they were applied in many cases where this was not the case. He wrote,

"It would appear that shortly before 1898 the process which leads to the correct value, of the probable errors of *optimum* statistics, was hit upon and found to agree with the probable errors of statistics found by the method of moments for *normal* curves and surfaces; without further enquiry it would appear to have been assumed that this process was valid in all cases, its directness and simplicity being peculiarly attractive. The mistake was at the time, perhaps, a natural one; but that it should have been discovered and corrected without revealing the inefficiency of the method of moments is a very remarkable circumstance" (Fisher, 1922a, page 329).

It is worth pointing out that the size of the correction Fisher noted was needed was not small. Fisher gave several examples of nonnormal members of Pearson's own family of curves where the lower bound of the efficiency of the moment-based estimates was zero. Since Fisher measured efficiency as a ratio of variances, this meant that the correction needed for Pearson's 1898 expressions for "probable errors" could be enormous—in fact arbitrarily large. The 1898 expressions were not larger than the actual probable errors, but there was little else that could be said. There was no finite limit to the amount they underestimated the actual probable errors.

The major error in the paper was due (as Fisher noted) to a conceptual confusion, a taking of the estimated frequency constants in part of the analysis in the place of the actual frequency constants. Pearson had run aground after encountering a need for a clear notion of a set of values for his frequency constants; he did not have a framework to encompass both estimates and targets of estimation. To some degree then, the Pearson–Filon error can be seen to be due to the lack of the notion of parametric families. Pearson and Filon used notation in this memoir suggestive of parametric families, but the lack of conceptual clarity led to a confused and ultimately erroneous analysis. Pearson thought of "frequency constants" as quantities such as moments, derivable from arbitrary density curves with the same meaning in all cases and with the sample moments as clearly leading to the best estimates.

Fisher's comments were apt; perhaps even too generous, although it is doubtful Pearson would have agreed with such an assessment. That the Pearson–Filon procedure can be shown to work for efficient estimates is a species of mathematical accident, albeit one that may have helped to deceive Pearson and probably produced overconfidence when it gave the results he knew should hold for the normal distribution case. The fact that the identities they claimed in general would work for many efficient estimates is mathematics that would have been foreign to Pearson and Filon, and unachievable without the full notion of parametric families.

From 1903 on, Pearson subtly distanced himself from the paper without ever calling attention to the errors, but he never repudiated it. In 1899 William F. Sheppard published a long study of "normal correlation" (Sheppard, 1899). Sheppard appears to have not seen the Pearson–Filon paper (at any rate he did not cite it), and a part of what he presented included probable errors for the frequency constants in the normal case, derived by methods different from Pearson–Filon. The methods he used were quite straightforward—writing estimates as linear functions of frequency counts (using a Taylor expansion if necessary), and then finding moments from the variances and covariances of the counts in ways that remain standard today.

The directness of Sheppard's methods must have appealed to Pearson. In a sequence of articles, all with the same title "On the probable errors of frequency constants" (Pearson, 1903, 1913, 1920), he presented what he called "simple proofs of the main propositions," all the while with the Pearson–Filon paper receding into the background. In 1903 he gave only a general reference to the 1898 treatment (as well as to Sheppard); in 1913 he only referred to the formulae in 1898 for the case of the normal correlation coefficient (where they were correct); in 1920 he did not cite the 1898 work at all. In the 1903 paper he included formulae based upon Sheppard's approach that were capable of being worked out for getting probable errors for estimates in five types of curves within the Pearson family, but only for methods of moments estimates.

The 1898 paper had a considerable impact upon statistical practice in making the use of probable errors available for the entire span of the new methodology including moment estimates. It could even be argued that the wrong, generally overoptimistic probable errors were better than none at all. And again, the paper had a significant impact upon Fisher. While preparing his 1922 memoir, Fisher clearly had Pearson and Filon before him, and his discussion of the asymptotic variance of maximum likelihood estimates (Fisher, 1922a, pages 328–329), involving the expansion of the density of a sample, reflects that. However, Fisher used the expansion in a different way, and operated under different assumptions. He began by assuming that the estimate tended to normality with large samples, and under that restriction and the assumption that the estimates maximized likelihood, Fisher used the expansion to show how the asymptotic variance could be found from the second derivative of the log density. Pearson and Filon had sketched a solution to a problem that was not the one they had embarked upon. But it was Fisher who recognized, with the conceptual apparatus of parametric families, that this sketch could lead to the solution of his own problem. Pearson the pioneer had laid a path that was insufficiently well-lit for his own travel, but it provided a brightly lit highway for Fisher.

## 4. PEARSON'S SECOND MAJOR ERROR

Pearson introduced the Chi-square test in 1900, and it has been widely celebrated as a great achievement in statistical methodology. In 1984 the editors of a popular science magazine selected it as one of twenty discoveries made during the twentieth century that have changed our lives (Hacking, 1984). Yet for all this celebration, virtually no historical mention of the paper is made by statisticians without adding damning words to the effect that Pearson erred in claiming, as we would now put it, that no correction in degrees of freedom need be made when parameters are estimated under the null hypothesis. Worse for Pearson's reputation, such accounts further note that the error stood uncorrected until it was sensed in 1915 by Greenwood and Yule and definitively corrected in 1922 and 1924 by Ronald Fisher, thus seemingly turning Pearson's landmark publication into Fisher's triumph over ignorance.

Pearson has had some defenders in this matter; some have even suggested that Pearson was right all along. For example, Karl's son Egon and George Barnard have separately advanced tentative (and I think half-hearted) statistical cases that might be made for proceeding as Pearson did (Pearson, 1938, page 30; Barnard, 1992). But a cold, clear-eyed look at the original 1900 paper shows that such excuses cannot be reconciled with Pearson's text. He did make an error, and a big, consequential one too.

The crucial passage from Pearson's 1900 article is on pages 165–166. Pearson considered a test of fit based upon a total of $N$ frequency counts from a sample independently distributed among $n+1$ groups or categories, with

$m = $ theoretical frequency [i.e., the expected frequency for the group in question],

$m_s = $ theoretical frequency deduced from data for the sample [i.e., expected frequency using the data to find the "best" value for the group],

$m' = $ observed frequency [for the group],

and with the total count $N = \sum m = \sum m_s = \sum m'$. Pearson recognized that the estimated theoretical frequency $m_s$ would typically differ from the theoretical frequency $m$, and he denoted that difference by $\mu$; that is, $\mu = m - m_s$. His analysis gave particular attention to the relative error, namely $\mu/m_s$, "which," he told us, "will, as a rule, be small."

The gist of Pearson's argument was to show that the Chi-square statistic based upon the theoretical frequencies, $\chi^2 = \sum \frac{(m'-m)^2}{m}$, is close to the Chi-square statistic based upon the estimated theoretical frequencies, $\chi_s^2 = \sum \frac{(m'-m_s)^2}{m_s}$; so close, in fact, that the discrepancy could for all practical purposes be ignored.[1]

Pearson had evidently expanded $h(m) = \frac{(m'-m)^2}{m} = \frac{(m'-m_s-\mu)^2}{m_s+\mu}$ in a Taylor series about $m_s$, discarded the terms of higher order than $(\mu/m_s)^2$, and then summed the results over the $n+1$ groups. Proceeding in this way, he would have found

$$h'(m) = -\frac{(m'^2 - m^2)}{m^2},$$

$$h''(m) = \frac{2m'^2}{m^3},$$

$$h'''(m) = -\frac{6m'^2}{m^4}, \ldots.$$

And so, since $\mu = m - m_s$,

$$h(m) = h(m_s) + \mu h'(m_s)$$
$$+ \frac{\mu^2}{2} h''(m_s) + \frac{\mu^3}{6} h'''(m_s) + \cdots$$

---

[1] Pearson again employed $S$ for $\sum$, and his argument is made harder than necessary to understand by two clear typographical errors. The typographical errors are an evident missing left parenthesis in the numerator of the second term on his first line of equations on page 165, and a missing $m_s$ in the denominator of the second term of the second line of equations [it reappeared, correctly, when this term was repeated two lines later; that equation is our equation (3) below].

$$= \frac{(m'-m_s)^2}{m_s} - \frac{\mu}{m_s}\frac{m'^2-m_s^2}{m_s}$$
$$+ \left(\frac{\mu}{m_s}\right)^2 \frac{m'^2}{m_s} - \left(\frac{\mu}{m_s}\right)^3 \frac{m'^2}{m_s} + \cdots$$
$$= \frac{(m'-m_s)^2}{m_s} - \frac{\mu}{m_s}\frac{m'^2-m_s^2}{m_s} + \left(\frac{\mu}{m_s}\right)^2 \frac{m'^2}{m_s},$$

dropping terms of higher order than $(\mu/m_s)^2$. Sum both sides over the $n+1$ groups and this is the expression Pearson arrives at:

(2)
$$\chi^2 = \chi_s^2 - \sum \left\{ \frac{\mu}{m_s} \frac{m'^2 - m_s^2}{m_s} \right\}$$
$$+ \sum \left\{ \left(\frac{\mu}{m_s}\right)^2 \frac{m'^2}{m_s} \right\}, \text{ and hence,}$$

(3)
$$\chi^2 - \chi_s^2 = -\sum \left\{ \frac{\mu}{m_s} \frac{m'^2 - m_s^2}{m_s} \right\}$$
$$+ \sum \left\{ \left(\frac{\mu}{m_s}\right)^2 \frac{m'^2}{m_s} \right\}.$$

The term $-m_s^2$ in the numerator of the first term on the right-hand side of (3) is superfluous when summed over groups since $\sum \mu = \sum m - \sum m_s = 0$, but it plays a role in Pearson's argument, which is no doubt why he left it in.

For future reference, I note that exactly the same result can be arrived at more simply by noting

$$\chi^2 = \sum \left\{ \frac{m'^2 - 2mm' + m^2}{m} \right\}$$
$$= \sum \left\{ \frac{m'^2}{m} \right\} - 2N + N = \sum \left\{ \frac{m'^2}{m} \right\} - N,$$

and similarly $\chi_s^2 = \sum \{ \frac{m'^2}{m_s} \} - N$; then

$$\chi^2 - \chi_s^2 = \sum \left\{ \frac{m'^2}{m} \right\} - \sum \left\{ \frac{m'^2}{m_s} \right\}$$
$$= \sum m'^2 \left( \frac{1}{m} - \frac{1}{m_s} \right).$$

If we then expand $m^{-1}$ as a function of $m$ about $m_s$ (again neglecting third- or higher order terms), we get

(4) $$\chi^2 - \chi_s^2 = -\sum \left\{ \frac{\mu}{m_s} \frac{m'^2}{m_s} \right\} + \sum \left\{ \left(\frac{\mu}{m_s}\right)^2 \frac{m'^2}{m_s} \right\}.$$

This agrees exactly with Pearson's expansion (3) when the superfluous term "$-m_s^2$" is dropped, as would have to be the case since the function being expanded $(\chi^2 - \chi_s^2)$ is the same in both cases.

It is not hard to show reasonably generally under the hypothesis of fit that the terms dropped, even when summed, are indeed with high probability negligible when $N$ is large $[O_P(N^{-1/2})]$. In order to see where Pearson was led astray, we must then look to the paragraph following his equations. Pearson's argument proceeded as follows: He recognized that the difference (3) between these two Chi-squares should be positive: the deviation of the observed counts from the theoretical counts should be greater than the same deviation if the theoretical counts are adjusted to fit the observed. He wished to argue that the difference (3) was not large. His argument was in two parts: (i) the first term on the right-hand side of (3) should be expected to be either negative (thus canceling out part of the second term) or at least very small; (ii) the second term was nonnegative of course, but it would be expected to be small in any case, because it involved for each summand the square of the relative error $\mu/m_s$, which Pearson had stated (page 164) "will, as a rule, be small," and much smaller still when squared. He gave no citation for this "rule," but two years earlier he had explicitly cited Gauss, Laplace and Poisson, among others, as sanctioning the dropping of terms involving the squares of errors thought to be small (Pearson and Filon, 1898, page 246). Presumably in stating this he assumed good estimates and ample data. He granted that in some cases where the fit was bad the deviations would be quite large, but then both Chi-squares would be large and the discrepancy between them unimportant.

There are two points to make about Pearson's argument. The first is that his analysis (i) of the first term may seem dubious to modern eyes, but it is not the source of the error. He noted that the first term will be positive only if the two terms multiplied ($\mu = m - m_s$ and $m'^2 - m_s^2$) are negatively correlated[2]; that is, if there was a tendency for the $m$'s to be ordered $m' > m_s > m$ or $m' < m_s < m$. He thought such a tendency "seems impossible," but this is unconvincing, at least under the null hypothesis of fit. Might we not then expect often to find $m' > m_s > m$ or $m' < m_s < m$, with $m_s$ a compromise between theory and observation? He might have had an alternative hypothesis in mind, where $m'$ would then tend to track the true theoretical expectation $m$, leaving the estimate $m_s$ (made under false assumptions) off to one side. Although Pearson's argument on point (i) can be questioned, his conclusion is correct. As Fisher would observe later, the first term is in fact zero (or nearly so) if the estimated $m$'s are chosen well (minimum Chi-square or maximum likelihood) due to the (near) orthogonality of $m - m_s$ and $m' - m_s$ in those cases (much like that of $\bar{X}$ and $X_i - \bar{X}$ for normal distributions).

In any event, it is part (ii) of his argument that is crucial, and that argument fails, and fails dramatically. The second term on the right-hand side of (3) should not be expected to be small under either null or alternative hypothesis. At this distance in time it may seem surprising that Pearson did not realize this. Already in 1938 his son Egon registered this surprise in a biographical memoir of his father (Pearson, 1938, page 30), when he noted that for any multinomial distribution, if Chi-square is computed with no parameter restrictions (so each theoretical value is estimated by the corresponding observed count and $m_s = m'$), then the fit with the estimated values is perfect. We would thus have $\chi^2 - \chi_s^2 = \chi^2 - 0$, while the right-hand side of (3) gives

$$-\sum\left\{\frac{\mu}{m_s}\frac{m'^2 - m_s^2}{m_s}\right\} + \sum\left\{\left(\frac{\mu}{m_s}\right)^2\frac{m'^2}{m_s}\right\}$$
$$= -0 + \sum\left\{\frac{(m-m')^2}{m'}\right\}.$$

In this extreme case the second term is asymptotically equivalent to the original Chi-square itself under the null hypothesis, and so it is certainly not negligible. The test of fit is not interesting here (we would say the degrees of freedom is zero), but it shows starkly the devastating effect estimated parameters can have upon the statistic, even when (as in Egon's example) the relative error itself ($\mu/m_s$) would be small $[O_P(N^{-1/2})]$. Why, Egon seemed to ask, would Karl have not seen this? Egon offered his father's possible "hurry in execution" as one explanation.

## 5. FISHER'S CORRECTION

Ironically, Pearson did consider a similar example in 1922 and rejected its relevance. In 1922 Fisher (1922b) published his first comment on the degrees of freedom issue, and at that time he dealt only with the case Greenwood and Yule had noticed, the case of $r \times c$ contingency tables. There, Fisher's argument was keyed to the way the linear relations with the marginal totals inhibited the estimated expectations under the null hypothesis, thus reducing the "degrees of freedom," a term Fisher introduced there. At that time, Fisher made no attempt to address the question for tests of fit more generally. Pearson immediately rebutted in *Biometrika*. The reply focused upon what Pearson thought (mistakenly) was a confusion between different sampling models (fixed totals or full multinomial sampling), and Pearson invoked the traditional custom of astronomers and others of substituting estimates with small standard errors without penalty in large samples. He thought Fisher had blundered and was offering an exclusively conditional analysis, given the estimated quantities. Pearson noted (1922, page 187) that if you estimated "the first $p-1$ moment-coefficients" a perfect fit would be obtained; he rejected such a conditional analysis as restricting the random sampling and antithetical to the question at issue. He did not see (and Fisher's exposition would have made it difficult for him to see) that in the contingency table setting the conditional and unconditional tests were the same.

---

[2]This would presumably be why he left the superfluous term "$-m_s^2$" in the expression.

In 1924 Fisher returned to address the more general question, and if we look at Fisher's treatment there, we see exactly where Pearson's argument about the second term of (3) failed, and exactly what he lacked for a successful treatment (Fisher, 1924). Writing in 1924, Fisher clearly had Pearson's paper in front of him. Fisher used slightly different notation,[3] but for ease of comparison I shall translate to Pearson's notation. Fisher's development was slightly streamlined in that Fisher did give the simpler expression for the difference of Chi-squares:

$$\chi^2 - \chi_s^2 = \sum m'^2 \left(\frac{1}{m} - \frac{1}{m_s}\right).$$

It is exactly this expression that Fisher expanded in a Taylor series, just as Pearson had done, but with one absolutely crucial difference. Fisher was now armed with his own recently introduced notion of a parametric family, and where Pearson had simply dealt with this as a function of $m$, Fisher had $m = m(\theta)$ and expanded as a function of $\theta$, not $m$. He found the same two terms Pearson had found, but expressed them differently:

$$\frac{1}{m} - \frac{1}{m_s}$$

---
[3] Fisher used $\chi', x, m'$ and $n$ where Pearson used $\chi_s, m', m_s$ and $N$.

$$= -\frac{1}{m_s^2}\frac{\partial m_s}{\partial \theta}\delta\theta$$
$$+ \left\{\frac{2}{m_s^3}\left(\frac{\partial m_s}{\partial \theta}\right)^2 - \frac{1}{m_s^2}\frac{\partial^2 m_s}{\partial \theta^2}\right\}\frac{(\delta\theta)^2}{2}$$
$$+ \text{higher-order terms}.$$

If this is multiplied by $m'^2$ and summed it gives

$$\chi^2 - \chi_s^2 = -\delta\theta \sum \left(\frac{m'^2}{m_s^2}\frac{\partial m_s}{\partial \theta}\right)$$
$$+ \frac{(\delta\theta)^2}{2}\sum\left\{\frac{2m'^2}{m_s^3}\left(\frac{\partial m_s}{\partial \theta}\right)^2 - \frac{m'^2}{m_s^2}\frac{\partial^2 m_s}{\partial \theta^2}\right\}.$$

Fisher was now able to see that if the minimum Chi-square estimate $\hat\theta$ is used, then his first term and Pearson's first term actually vanished (since then the first summation is exactly $\frac{d}{d\theta}\chi^2|_{\theta=\hat\theta}=0$), and he knew already that the same would be true asymptotically for the maximum likelihood estimate or any other efficient estimate of $\theta$. He then replaced $m'/m_s$ by unity (its asymptotic value) to get

$$\chi^2 - \chi_s^2 = (\delta\theta)^2 \sum\left\{\frac{m'^2}{m_s^3}\left(\frac{\partial m_s}{\partial \theta}\right)^2 - \frac{m'^2}{2m_s^2}\frac{\partial^2 m_s}{\partial \theta^2}\right\}$$
$$\approx (\delta\theta)^2 \sum\left\{\frac{1}{m_s}\left(\frac{\partial m_s}{\partial \theta}\right)^2 - \frac{1}{2}\frac{\partial^2 m_s}{\partial \theta^2}\right\}$$
$$= (\delta\theta)^2 \sum\left\{\frac{1}{m_s}\left(\frac{\partial m_s}{\partial \theta}\right)^2\right\}.$$

The last step used the fact that $\sum\{\frac{\partial^2 m_s}{\partial \theta^2}\} = \frac{\partial^2}{\partial \theta^2} \cdot \sum m_s = \frac{\partial^2}{\partial \theta^2} N = 0$. Based upon his own 1922 paper, he now noted that $\sum\{\frac{1}{m_s}(\frac{\partial m_s}{\partial \theta})^2\}$ would, in the case of a single estimated parameter $\theta$, estimate (and approximate asymptotically) the reciprocal of the variance of any efficient estimate. This would give in modern notation $\chi^2 - \chi_s^2 = \frac{(\hat\theta-\theta)^2}{\sigma^2(\hat\theta)}$. This difference then was asymptotically equivalent to the square of a standard normal random variable. The degree of freedom that is lost by estimation became clearly visible.

There are two views that may be taken of this. One I have already mentioned: that the alchemist Fisher's concept of a parametric family had turned Pearson's base expressions into statistical gold. Posterity has used this to diminish Pearson's reputation—how could he have missed such a simple and (now) obvious step? But there is another, to me more persuasive view. For over 20 years that step was anything but obvious. Pearson's perceptive student G.

Udny Yule initially accepted the 1900 rule, for example using 8 rather than 4 degrees of freedom for a Chi-square test of a $3 \times 3$ contingency table in Yule (1906, page 349). Only in 1915, after years of experience, did Greenwood and Yule (1915) bring the puzzle to wider notice, and even then neither they nor anyone else had a clear view of the source of the problem. And so it stood until Fisher.

Even with Fisher's work before us, we must marvel at how far Pearson had gone. He had lacked only one ingredient—parametric families—but what he had managed to do was to identify the issue and present it in such a clear way that when Fisher combined Pearson's 1900 development with the deceptively simple idea of parametric families, the solution must have sprung to mind nearly immediately. It took Fisher's genius to answer the question, but he would scarcely have been in a position to do so without the path-breaking formulation of Pearson the pioneer.

It is not anachronistic to see Pearson as erring in 1900. Even without the notion of parametric families he could have seen a discrepancy without seeing a resolution, just as Greenwood and Yule did, when they found the Chi-square test for $2 \times 2$ tables gave results inconsistent with a comparison of the two columns as binomial counts. Pearson erred, but the error led to Fisher's discovery of degrees of freedom. Pearson had not only solved the great problem of testing multinomial goodness of fit against all alternatives, he had also isolated and formulated another great problem in terms that two decades later permitted another genius, armed with his own major discovery, an easy solution.

## 6. CONCLUSION

The errors Pearson made did not go undetected because they were small; to the contrary, they were large and of potentially large practical consequence. For example, in Pearson and Filon's own numerical example for a Type III or Gamma density (Pearson and Filon, 1898, pages 279–280), the probable error given for the shape parameter $p$ is only about a fifth of what it should have been (Fisher, 1922a, page 336). If the curves being fit by the method of moments had been closer to the normal shape, the errors would have been smaller, but if not, there was no finite bound on how far off they could be. For Chi-squares for $2 \times 2$ tables Pearson would give 3 rather than 1 degree of freedom; for $3 \times 3$ tables he would give 8 rather than 4 degrees of freedom. In these and other examples the effect upon inferences could be devastating.

Not only were the errors Pearson made not easily discovered; even after they were pointed out in 1922 they were not widely understood. In 1924, a *Handbook of Mathematical Statistics* was published, prepared under the auspices of the U.S. National Research Council (Rietz, 1924). The Editor-in-Chief was H. L. Rietz, and major contributions were made by Harvard University Professor E. V. Huntington and University of Michigan Professor H. C. Carver (later the founding editor of the *Annals of Mathematical Statistics*). Carver cited the Pearson–Filon paper without any indication he saw the problem with it (page 95). Fisher's (1922b) first correction to the degrees of freedom for contingency tables was briefly cited without comment by Rietz, but Rietz (with evident approval) also gave in more detail Pearson's argument that no correction for estimating expected values was needed (pages 80–81). Elsewhere in the volume, Huntington wrote warmly of the method of moments, and nowhere was Fisher's magnum opus of 1922 referred to. Even in England understanding was slow. By 1938 Egon Pearson had conceded the degrees of freedom issue, but he seemed to have not accepted Fisher's point about Pearson and Filon (Pearson, 1938, pages 28–29).

Both Pearson and Fisher were giants in our history; despite their lack of mutual appreciation we cannot imagine modern statistics without both. Pearson's errors were substantial and not to be glossed over, but they should not obscure the even greater achievements they accompanied. Pearson had a giant ambition and the energy to realize it. He sought to create a whole new statistical system, and for a time succeeded. He did not have a mathematical mind equal to Fisher's, and he became mired in and never escaped from an incompletely developed conceptual apparatus that was not equal to the full task at hand. But he took statistics to a higher level nonetheless. If Pearson could never come to admit some failures, it was surely due to a stubbornness that even he recognized in himself. In the Preface for the Second Edition of *The Grammar of Science* (1900b), Pearson wrote,

> "If I have not paid greater attention to my numerous critics, it is not that I have failed to study them; it is simply that I have remained—obstinately it may be—convinced that the views expressed are, relatively to our present state of knowledge, substantially correct" (Pearson, 1900b, page ix).

So it was with his statistical work as well.

Pearson's impact upon Fisher may in the end stand as one of his greater achievements. Pearson had no student more diligent than Fisher, despite their differences. When in 1945 Fisher wrote an ill-fated biographical account of Pearson for the *Dictionary of National Biography* (rejected by the *Dictionary* and not published until by A. W. F. Edwards, in 1994), he wrote to the editor that he had made a "lifelong study of Pearson's writings." Fisher further stated, "I have during the last 35 years at various times had occasion to look at probably all of [Pearson's fundamental statistical memoirs] and at the immense output which was published in *Biometrika*." It was from reading Pearson's work and Pearson's journal that Fisher's interest in statistics developed in the way it did, and in the case of the two examples discussed here, the effect of the Pearsonian blueprint could scarcely be more evident. Fisher saw Pearson clearly, warts and all, and while he did not acknowledge the extent of his debt to Pearson, its extent is clear to other, less involved readers. As in Newton's famous statement, Fisher stood on the shoulders of a giant (Merton, 1965).

Porter's recent biography (2004) is illuminating on Pearson's pre-statistical life. Eisenhart (1974) remains the most complete discussion of K. P.'s statistical work. For other discussion relating to this early work see Aldrich (1997), Hald (1998), Magnello (1996, 1998). On Chi-square see in particular Fienberg (1980), Hacking (1984), Plackett (1983) and Stigler (1999, Chapter 19). For other aspects of the Pearson–Fisher relationship see Stigler (2005, 2007a, 2007b). Pearson himself returned to that topic of Chi-square frequently, including Pearson (1915, 1922, 1923, 1932), most of these under the instigation of Fisher.